\documentclass[10pt,a4paper,twocolumn,showpacs,showkeys,lengthcheck,nobibnotes,nofootinbib]{revtex4}

\usepackage{amsfonts}
\usepackage{amsmath}
\usepackage{amssymb}
\usepackage{fancyhdr}
\usepackage{natbib}

\usepackage[compact]{titlesec}
\titlespacing{\section}{0pt}{10pt}{3pt}
\titlespacing{\subsection}{0pt}{10pt}{0pt}
\titlespacing{\subsubsection}{0pt}{10pt}{0pt}

\begin{document}

\title{
\textbf{The Fock-Kemmer approach to precursor \\ 
shock waves in relativistic field theory}}
\author{Rawand Abdullah$^{1}$ and George Jaroszkiewicz$^{2}$}
\affiliation{\small $^{1}$University of Sulaimani, School of Science Educations,\\ 
Department of Physics, City Campus, Zanko St., As-Sulaimaniah, Kurdistan Region, Iraq \\
$^{2}$george.jaroszkiewicz@nottingham.ac.uk }
\author{\textbf{\today}}
\keywords{shock waves, field theory, characteristics, propagators}
\pacs{52.35.Tc,03.50.-z,11.15.-q,03.50.De,52.40.Db,02.50.Ng}

\begin{abstract}
\begin{center}
\textbf{Abstract}
\end{center} \vspace{-10pt}
We use distribution theory (generalized functions) to extend and justify the Fock-Kemmer approach to the propagation of precursor shock wave discontinuities in 
classical and quantum field theory. We apply lightcone causality arguments to propose that shock waves singularities in non-linear classical field theories and in Maxwell's equations for responsive media require a form of classical renormalization analogous to Wilson operator product expansions in quantum field theories.  
\end{abstract}

\maketitle


\section{Introduction}

Precursor shock waves are physically observable, causal discontinuities in
classical and quantum fields propagating through regions of relativistic
spacetime. By this is meant a phenomenon such that, before a certain time $%
t_{0}$ in some laboratory or equivalent, an array of signal detectors is in
its no-signal state, but after that time, geometrically recognizable
patterns of signals will be found to have been triggered in that array.

This subject has become topical on account of recent reports of the
observation of gravitational shock waves by the LIGO and Virgo science
collaborations \citep{ABBOTT+ETAL-2016}, but it has a long and important
history. The advent of special relativity raised a fundamental question: are
Maxwell's equations consistent with the lightcone veto on superluminal
signal propagation? The issue was tackled in 1914 by Sommerfeld and
subsequently by Brillouin \citep{STRATTON-1941}, with much theoretical and
empirical interest since then. Empirically important examples of great
interest are Cerenkov radiation, or the shock waves in media produced by
superluminal charged particles and the Askaryan effect, or the shock waves
in media produced by superluminal neutral particles. In addition to these
and the above mention gravity wave detection, two greatly discussed cases
involving neutrinos illustrate the critical importance of this subject:
\noindent $i)$ ultra-relativistic neutrino signals from supernova Sn 1987A
allowed limits to be placed on neutrino masses and the number of neutrino
flavours \citep{NOMOTO+SHIGEYAMA-1987}, and $ii)$ in 2011, subsequently
refuted reports from the OPERA experiment of superluminal neutrino signals
threatened to undermine special relativity.

In this article our focus is not on the detectors per se but on the
classical or quantum fields through which discontinuities are carried. Our
strategy is not to solve the relevant differential equations but to use
causality principles and the information encoded in those equations directly
to discuss the speed of precursor shock waves. In this we have been guided
by the approach of Fock in his discussion of electromagnetic signals in
general relativity \citep{FOCK-1964}. This approach stands in contrast with
conventional explicit methods, such as that of Summerfeld and Brillouin and
to the numerical simulation of relativistic shock waves \cite%
{LEMOINE+PELLETIER-2011}.

It is generally assumed that all physical effects propagate at speeds
limited by the light cone structure of relativistic spacetime, but important
questions remain to be answered. For instance, two important speeds are
generally discussed in quantum wave mechanics. One is the \emph{phase}
velocity\footnote{%
Actually a speed, but the term \emph{phase velocity} sounds better.} $w$
conventionally associated with de Broglie waves and the other is the
particle speed $v$, frequently referred to as \emph{group velocity}. These
speeds satisfy the de Broglie relation $wv=c^{2}$, where $c$ is the speed of
light. But for massive particles, neither of these speeds is equal to $c$
and since $v$ has to be less than $c$ for reasons of causality,we deduce
that $w$ is superluminal. The conventional interpretation of this is that $w$
is associated with correlations. Two correlated events can be observed on a
hyperplane of simultaneity (even in Newtonian mechanics), thereby giving the
impression of superluminal speeds \citep{SCARANI-2000}. However, that does
not prove any causal connection, even in classical physics. Correlations
have everything to do with the context of observation, including how the
correlated events were set up in the first place. Other questions concern
the causality structure of the Feynman propagator in relativistic quantum
field theory, and the propagation of higher spin fields, such as the
Rarita-Schwinger field \citep{RARITA-SCHWINGER-1941}, as these appear to
involve superluminal speeds\citep{VELO+ZWANZIGER-1969A}. We shall comment on
all of these issues.

In this article we focus on a third speed involved in field theory, that is,
the speed of propagation of shock waves. Any analysis of shock waves
requires a careful interplay between reductionist and emergent concepts. On
the one hand, field equations are generally derived from the reductionist
principles of Lagrangian mechanics. On the other hand, shock waves are large
scale, emergent processes highly sensitive to the details of those field
equations, the non-local initial conditions setting off those shock waves,
the laws of causality, and the protocols of observation.

In the standard approach to field theory, the mathematics is usually
discussed from the perspective of an exophysical observer standing outside
of some region of spacetime, monitoring the behaviour of a system under
observation in that region. The observer usually enters the picture in only
two places: the first is where they initialize the equations of motion
describing the system and the second is then when they observe the final
state of that system. Certainly that is the way quantum theory is normally
discussed when it is applied to scattering processes. In any discussion of
shock waves, however, the situation becomes more complicated. Now the role
of the observer becomes more intermingled with the dynamical evolution of
the system under observation, requiring more care and detail in the analysis.

Throughout this paper, the term \emph{suitably arbitrary} means arbitrary
provided certain conditions such as differentiability are met. We set $%
c=\hslash =1$ and work in a standard Minkowski spacetime inertial frame with
metric tensor components $(1,-1,-1,-1)$ down the main diagonal and zero
everywhere else.

\ In the next section we discuss a simplified model that serves as a
template for all discussions in this paper.

\section{First order linear PDE}

\label{FIRSTORDER}In $1+3$ spacetime and relative to an inertial frame with
coordinates $(x^{0}\equiv t,\boldsymbol{x})$, consider the field equation%
\begin{equation}
i\dot{\phi}+i\boldsymbol{a} \boldsymbol{\cdot } \nabla \phi -m\phi \underset{%
c}{=}0,  \label{FKW-102-1}
\end{equation}%
where $\phi $ is a real or complex scalar field, $\boldsymbol{a}$ is a
non-zero, constant real $3$-vector, $m$ is a real constant, $\dot{\phi}%
\equiv \partial \phi (t,\boldsymbol{x})/\partial t,$ and $\underset{c}{=}$
denotes an equality holding only for solutions to (\ref{FKW-102-1}). We
investigate the possibility of finding shock wave solutions to equation (\ref%
{FKW-102-1}) in five nominally different approaches. The first three
approaches require us to solve the equation in one way or another. The merit
of the fourth approach, which is based on the work of Fock \citep{FOCK-1964}
and on Kemmer's notation \citep{KEMMER-1971}, is that it is easier in this
respect: we do not need to solve the differential equation but draw our
conclusions based on the structure of the differential equations themselves
and on the logic of observation and causality as it applies to shock waves.
This approach is similar in spirit to standard discussions of
characteristics given in \citep{COURANT+HILBERT-1953B} and applied by \citep%
{VELO+ZWANZIGER-1969B}. Towards the end of this paper we shall introduce a
fifth approach, based on distribution theory, that justifies the heuristic
approach of Fock.

\subsection{The Fourier transform approach}

Fourier transforming equation (\ref{FKW-102-1}) with respect to the spatial
coordinates, solving the transformed equation in transform space, and then
inverting back gives the general solution%
\begin{equation}
\phi (t,\boldsymbol{x})\underset{c}{=}e^{-imt}\Phi (\boldsymbol{x}-%
\boldsymbol{a}t),  \label{FKW-102-4}
\end{equation}%
where $\Phi $ is a suitably arbitrary function of the variable $\boldsymbol{z%
}\equiv \boldsymbol{x}-\boldsymbol{a}t$. Discontinuities can be embedded in
the shape function $\Phi $. For example, a typical plane wave Tsunami type
of solution will be of the form $\Phi (\boldsymbol{z})=f(\boldsymbol{z}%
)\theta (t-\boldsymbol{n}\boldsymbol{\cdot} \boldsymbol{x})$, where $%
\boldsymbol{n}\equiv \boldsymbol{b}/(\boldsymbol{a\cdot b})$ for any vector $%
\boldsymbol{b}$ such that $\boldsymbol{a} \boldsymbol{\cdot} \boldsymbol{b}%
\neq 0$, $f$ is suitably arbitrary, and $\theta $ is the Heaviside step
function.

\subsection{The method of characteristics}

In this approach we first rewrite (\ref{FKW-102-1}) in matrix form: 
\begin{equation}
i[1,\boldsymbol{a}^{T}]%
\begin{bmatrix}
\partial _{t} \\ 
\nabla%
\end{bmatrix}%
\phi (t,\boldsymbol{x})-m\phi (t,\boldsymbol{x})\underset{c}{=}0,
\label{FKW-111}
\end{equation}%
where superscript $T$ denotes transpose. Next, we make the passive
linear-inhomogeneous coordinate transformation%
\begin{equation}
\begin{bmatrix}
t^{\prime } \\ 
\boldsymbol{x}^{\prime }%
\end{bmatrix}%
=%
\begin{bmatrix}
\alpha & \boldsymbol{\beta }^{T} \\ 
\boldsymbol{\gamma } & \delta I_{3}%
\end{bmatrix}%
\begin{bmatrix}
t \\ 
\boldsymbol{x}%
\end{bmatrix}%
+%
\begin{bmatrix}
s \\ 
\boldsymbol{r}%
\end{bmatrix}%
,  \label{FKW-222}
\end{equation}%
where $\alpha ,\delta $ and $s$ are real constants, $\boldsymbol{\beta }$, $%
\boldsymbol{\gamma }$ and $\boldsymbol{r}$ are real column three-vectors,
and $I_{3}$ is the $3\times 3$ identity matrix. This transformation is
invertible provided $(\alpha \delta -\boldsymbol{\beta }\boldsymbol{\cdot} 
\boldsymbol{\gamma })\delta ^{2}\neq 0$, which we assume. Given that $\phi $
is a scalar field and defining $\phi ^{\prime }(t^{\prime },\boldsymbol{x}%
^{\prime })\equiv \phi (t,\boldsymbol{x})$, (\ref{FKW-111}) becomes%
\begin{equation}
i\left\{ \left( \alpha +\boldsymbol{a}\boldsymbol{\cdot} \boldsymbol{\beta }%
\right) \partial _{t^{\prime }}+\left( \boldsymbol{\gamma }+\delta 
\boldsymbol{a}\right) \boldsymbol{\cdot} \nabla ^{\prime }\right\} \phi
^{\prime }(t^{\prime },\boldsymbol{x}^{\prime })-m\phi ^{\prime }(t^{\prime
},\boldsymbol{x}^{\prime })\underset{c}{=}0.  \label{FKW-444}
\end{equation}%
We now take advantage of the fact that the various constants in
transformation (\ref{FKW-222}) are suitably arbitrary. We choose to set%
\begin{equation}
\boldsymbol{\gamma }+\delta \boldsymbol{a}=\boldsymbol{0},
\label{FKW-102-10}
\end{equation}%
and then (\ref{FKW-444}) becomes%
\begin{equation}
i\left( \alpha +\boldsymbol{a}\boldsymbol{\cdot} \boldsymbol{\beta }\right)
\partial _{t^{\prime }}\phi ^{\prime }(t^{\prime },\boldsymbol{x}^{\prime
})-m\phi ^{\prime }(t^{\prime },\boldsymbol{x}^{\prime })\underset{c}{=}0.
\label{FKW-333}
\end{equation}%
Assuming $\alpha +\boldsymbol{a}\boldsymbol{\cdot} \boldsymbol{\beta }\neq 0$%
, the general solution to (\ref{FKW-333}) is%
\begin{equation}
\phi ^{\prime }(t^{\prime },\boldsymbol{x}^{\prime })=U(\boldsymbol{x}%
^{\prime })\exp \left\{ -\frac{imt^{\prime }}{\alpha +\boldsymbol{a}%
\boldsymbol{\cdot} \boldsymbol{\beta }}\right\} ,
\end{equation}%
where $U$ is suitably arbitrary. Transforming back to the original
coordinates and using (\ref{FKW-102-10}) we get%
\begin{equation}
\phi (t,\boldsymbol{x})=U(-\delta \boldsymbol{a}t+\delta \boldsymbol{x+r}%
)\exp \left\{ -\frac{im(\alpha t+\boldsymbol{\beta }\boldsymbol{\cdot} 
\boldsymbol{x}+s)}{\alpha +\boldsymbol{a}\boldsymbol{\cdot} \boldsymbol{%
\beta }}\right\} ,
\end{equation}%
which is equivalent to (\ref{FKW-102-4}), the solution found using the
Fourier transform method.

\subsection{The Schwinger-Pauli-Jordan function method}

This is perhaps the most powerful method in standard free field theory, as
it explicitly solves the initial value problem in Lorentz-signature
spacetimes (critical to a satisfactory physical interpretation of what is
going on) as well as explicitly revealing the causal singularity structure
that ultimately underpins the propagation of shock waves.

Solutions to (\ref{FKW-102-1}) are assumed to have the form%
\begin{equation}
\phi (t,\boldsymbol{x})\underset{c}{=}\int d^{3}\boldsymbol{y}G^{(+)}(t,%
\boldsymbol{x}-\boldsymbol{y})\eta (\boldsymbol{y})  \label{FKW-1167}
\end{equation}%
for $t>0$. Here $\{\eta (\boldsymbol{y}):y\in \mathbb{R}^{3}\}$ represents
the initial data, that is, the field values distributed over the spacelike
hypersurface at initial laboratory time $t=0$. Taking into account the fact
that the observer necessarily exists before the shock wave is initiated, the
Schwinger-Pauli-Jordan (SPJ) function $G^{(+)}$ is taken here to be a
distribution over the spacetime $(-\infty ,\infty )\times \mathbb{R}^{3}$
with the following properties:

\begin{enumerate}
\item $G^{(+)}(t,\boldsymbol{x})=0,\ \ \ t<0;$

\item $\left( i\partial _{t}+i\boldsymbol{a}\boldsymbol{\cdot} \nabla
_{x}-m\right) G^{(+)}(t,\boldsymbol{x})=0,\ \ \ \ \ t>0$,

\item $\lim_{t\rightarrow t_{0}+}G^{(+)}(t,\boldsymbol{x})=\delta ^{3}(%
\boldsymbol{x})$.
\end{enumerate}

Given these conditions, the SPJ function is readily found to be%
\begin{equation}
G^{(+)}(t,\boldsymbol{x})=\theta (t)e^{-imt}\delta ^{3}(\boldsymbol{a}t-%
\boldsymbol{x}),
\end{equation}%
ignoring any inessential $\delta (t)$ contribution. The interpretation of
this solution is that it encodes the shock wave that would be propagated
throughout future spacetime from a point event disturbance at the origin of
space and time coordinates. The Heaviside function has been inserted here by
hand to reinforce the classical causality condition that the field $\phi $
cannot exist before it is created at initial time $t_{0}$. As a distribution
over all spacetime, $G^{(+)}$ does not satisfy the original homogeneous
equation of motion but does satisfy the inhomogeneous equation%
\begin{equation}
\left( i\partial _{t}+i\boldsymbol{a}\boldsymbol{\cdot} \nabla _{x}-m\right)
G^{(+)}(t,\boldsymbol{x})=i\delta (t)\delta ^{3}(\boldsymbol{a}t - 
\boldsymbol{x}),
\end{equation}%
reflecting the creation of a point source at time zero.

Since the original wave equation (\ref{FKW-102-1}) is linear, shock waves
from different point sources would not interact with each other, but would
superpose. Therefore, the combined effect of a collection of such events
distributed over some spacelike hypersurface is given by integrals such as (%
\ref{FKW-1167}).

This also applies if for instance the initial shock wave is generated in
some finite four-dimensional region $V$ of spacetime. Assuming no dynamical
interaction between fields created at different times, then the general
solution outside of this region will be given by%
\begin{equation}
\phi (t,\boldsymbol{x})\underset{c}{=}\int_{V}dt_{0}d^{3}\boldsymbol{y}%
G^{(+)}(t-t_{0},\boldsymbol{x}-\boldsymbol{y})\varrho (t_{0},\boldsymbol{y}),
\end{equation}%
where $\varrho (t_{0},\boldsymbol{y})$ represents a spacetime density of
source events and the Heaviside function in $G^{(+)}$ ensures classical
causality is obeyed at all times. By this we mean that in this scenario,
every point source event can influence events only in its own relative
future. A similar, implicit assumption is made in Schwinger's source theory 
\citep{SCHWINGER-1969}.

Inside the region $V$, the field $\phi (t,\boldsymbol{x})$ satisfies the
inhomogeneous equation%
\begin{equation}
\left( i\partial _{t}+i\boldsymbol{a}\boldsymbol{\cdot} \nabla _{x}-m\right)
\phi (t,\boldsymbol{x})=i\varrho (t,\boldsymbol{x}),
\end{equation}%
which could be used to model the creation of a shock wave.

\subsection{The Fock-Kemmer approach}

The Fock-Kemmer approach to shock wave analysis is useful and economical
because it does not require any solution \emph{per se} of the differential
equations involved for conclusions about shock waves to be reached. Before
we can discuss the method, however, we need to introduce the concepts of 
\emph{Fock subsurface}, \emph{Fock flow}, \emph{subsurface normal velocity},
and \emph{Kemmer bracket}.

In the following, we assume we are an exophysical observer looking in over a
region $\mathcal{R}$ of $1+3$ dimensional spacetime, using a coordinate
patch $P(t,\boldsymbol{x})$ covering $\mathcal{R}$, such that the coordinate 
$t \in [0,T]$ represents observer time indexing a spacelike foliation of $%
\mathcal{R}$.

\subsubsection{\textbf{Fock subsurfaces}}

A \emph{Fock subsurface} $F_{t}$ at time $t$ is the set of points in $%
\mathcal{R}$ satisfying the condition 
\begin{equation}
F_{t}\equiv \{\boldsymbol{x}:F(\boldsymbol{x})=t, (t,\boldsymbol{x)}\in 
\mathcal{R\}},
\end{equation}%
where $F$ is some differentiable function of spatial coordinates only. Fock
subsurface functions are in general defined contextually by observers, such
as when torches and particle beams are switched on, or by natural causes
such as underwater avalanches or the collision of two black holes as
recently reported \citep{ABBOTT+ETAL-2016}. For example, in Newtonian
space-time, a spherical pulse of light generated at the origin of space-time
coordinates is subsequently distributed over a Fock surface defined by $%
\sqrt{\boldsymbol{x}\boldsymbol{\cdot} \boldsymbol{x}}=t$.

\subsubsection{\textbf{Fock flows}}

A \emph{Fock flow} $\mathcal{F}[F]$ is a family of Fock subsurfaces in $%
\mathcal{R}$ indexed by the observer's time $t $, that is, a family of
two-dimensional surfaces defined by the set of equations 
\begin{equation}
\mathcal{F}[F]\equiv \{F_{t}:t\in \lbrack t_{i},t_{f}]\},
\end{equation}%
where $F$ is a Fock subsurface function and $t_{i}<t_{f}$.

\subsubsection{\textbf{Subsurface normal velocity}}

The \emph{gradient} $\nabla F_{P}$ at a point $P$ on a Fock surface $F_{t}$
denotes the usual set of Cartesian spatial coordinate partial derivatives of 
$F$ evaluated at $P$. Given a Fock flow $\mathcal{F}[F]$, by considering a
point $P$ on the Fock subsurface $F_{t},$ projecting that point normally to
that subsurface so as to intersect the Fock subsurface F$_{t+\delta t}$, and
then taking the appropriate limit $\delta t\rightarrow 0$, it is
straightforward to establish that the ``velocity'' $\boldsymbol{w}_{P}$ at a
point $P$ on a given Fock subsurface is given by%
\begin{equation}
\boldsymbol{w}_{P}=\left( \nabla F_{P}\right) ^{-2} \nabla F_{P}.
\end{equation}%
This requires the gradient $\nabla F$ not to vanish at $P$. This velocity
will be referred to as the \emph{subsurface normal velocity} at $P$. Its
magnitude is the subsurface normal speed $w_{P}$ and is given by%
\begin{equation}
w_{P}=|\nabla F|^{-1}.  \label{FKW-speed}
\end{equation}

\subsubsection{\textbf{Kemmer brackets}}

Kemmer brackets were introduced \citep{KEMMER-1971} as a powerful notational
way to discuss Fock's shock wave analysis \citep{FOCK-1964}. Given a
propagating field $\phi $ and a Fock flow $\mathcal{F}[F]$, the Kemmer
bracket $[\phi ]^{F}$ of the field $\phi $ relative to $\mathcal{F}$ is
defined by%
\begin{equation}
\lbrack \phi ]^{F}(\boldsymbol{x})\equiv \phi (F(\boldsymbol{x}),\boldsymbol{%
x}).
\end{equation}%
A Kemmer bracket is a function of spatial coordinates only.

Here and elsewhere we shall make extensive use of the \emph{Fock-Kemmer
identity} 
\begin{equation}
\nabla [\phi ]^{F}=\nabla F\ [\dot{\phi}]^{F}+[\nabla \phi ]^{F},
\label{FKW-ID}
\end{equation}%
where $\lbrack \dot{\phi}]^{F}(\boldsymbol{x})\equiv \left. \partial
_{t}\phi (t,\boldsymbol{x})\right\vert _{t=F(\boldsymbol{x})}$ and $[\nabla
\phi ]^{F}(\boldsymbol{x}) \equiv \left. \nabla \phi (t,\boldsymbol{x}%
)\right\vert _{t=F(\boldsymbol{x})}$. We may apply the Fock-Kemmer identity
to derivatives of the field $\phi $, giving for example $\nabla [\dot{\phi}%
]^{F}=[\ddot{\phi}]^{F}\nabla F\ +[\nabla \dot{\phi}]^{F}$, and so on.

\subsubsection{\textbf{Application to equation (\protect\ref{FKW-102-1})}}

Considering the vector $\boldsymbol{a}$ in the original equation of motion (%
\ref{FKW-102-1}), the Fock-Kemmer identity gives 
\begin{equation}
\boldsymbol{a}\boldsymbol{\cdot} \nabla \lbrack \phi ]^{F}=\boldsymbol{a}%
\boldsymbol{\cdot} \nabla F[\dot{\phi}]^{F}+[\boldsymbol{a}\boldsymbol{\cdot}
\nabla \phi ]^{F}.  \label{FKW-102-3}
\end{equation}
On the other hand, applying the Kemmer bracket to the equation of motion (%
\ref{FKW-102-1}) directly gives%
\begin{equation}
i[\dot{\phi}]^{F}+i[\boldsymbol{a}\boldsymbol{\cdot} \nabla \phi
]^{F}-m[\phi ]^{F}\underset{c}{=}0.  \label{FKW-102-2}
\end{equation}%
Using (\ref{FKW-102-2}) in (\ref{FKW-102-3}) then gives%
\begin{equation}
(\boldsymbol{a}\boldsymbol{\cdot} \nabla F-1)[\dot{\phi}]^{F} \underset{c}{=}
\boldsymbol{a}\boldsymbol{\cdot} \nabla \lbrack \phi ]^{F}+im[\phi ]^{F}.
\label{FKW-102-5}
\end{equation}%
It is straightforward to verify that on the Fock subsurface $F(\boldsymbol{x}%
)=t$, the solution (\ref{FKW-102-4}) satisfies (\ref{FKW-102-5}), where now%
\begin{equation}
\lbrack \phi ]^{F}(\boldsymbol{x})\equiv e^{-imF(\boldsymbol{x})}\Phi (%
\boldsymbol{x}-\boldsymbol{a}F(\boldsymbol{x})).  \label{FKW-102-5555}
\end{equation}

\subsubsection{\textbf{The Fock shock wave condition}}

In the above, the Fock subsurface $F$ function is suitably arbitrary. Now
consider a specific choice, written $F = W$, representing a shock wave of
discontinuity. Fock's heuristic argument \citep{FOCK-1964} is that on such a
shock wave, it should not be possible to work out the Kemmer bracket of $[%
\dot{\phi}]^{W}$ from a knowledge of $[\phi ]^{W}$ or its derivatives such
as $\boldsymbol{a}\boldsymbol{\cdot} \nabla [\phi ]^{W}$. The constructs $%
[\phi ]^{W}$ and $\boldsymbol{a}\boldsymbol{\cdot} \nabla [\phi ]^{W}$
depend on initial data available in principle to the observer whilst $[\dot{%
\phi}]^{W}$ represents data that is causally unavailable. We shall call this
chain of reasoning \emph{Fock's argument}. Our distribution theory approach
in \S \ref{DISTFIELDAPP} fully justifies Fock's heuristic argument.

Given the Fock argument, then the conclusion from (\ref{FKW-102-5}) is that
the coefficient of $[\boldsymbol{a} \boldsymbol{\cdot }\nabla \phi ]^{W}$ on
the left-hand side of (\ref{FKW-102-5}) must vanish when $F=W$, that is, on
a surface of discontinuity. We deduce that a shock wave must satisfy the
equation 
\begin{equation}
\boldsymbol{a} \boldsymbol{\cdot} \nabla W=1.  \label{FKW-102-6}
\end{equation}%
This also means that the right-hand side of (\ref{FKW-102-5}) must vanish on
such a surface also, giving the condition%
\begin{equation}
\boldsymbol{a}\boldsymbol{\cdot} \nabla \lbrack \phi ]^{W}+im[\phi ]^{W}%
\underset{c}{=}0.  \label{FKW-102-7}
\end{equation}%
It is readily confirmed that (\ref{FKW-102-5555}) does indeed satisfy (\ref%
{FKW-102-7}) when $W$ satisfies the shock wave condition (\ref{FKW-102-6}).

\subsubsection{\textbf{Interpretation}}

To get some understanding of these results, we can without loss of
generality take $\boldsymbol{a}=(a,0,0)$ where $a>0.$ Then (\ref{FKW-102-6})
reduces to%
\begin{equation}
a\partial _{x}W(x,y,z)=1.
\end{equation}%
This equation has general solution%
\begin{equation}
W(x,y,z)=\frac{x}{a}+U(y,z),\ \ \ \ \ a\neq 0,  \label{FKW-102-8}
\end{equation}%
where $U$ is suitably arbitrary. Assuming the solution is of the form (\ref%
{FKW-102-5555}) we have%
\begin{equation}
\lbrack \phi ]^{W}(\boldsymbol{x})=e^{-imx/a-imU(y,z)}\Phi (U(y,z),-y,-z).
\end{equation}%
Then we readily find that condition (\ref{FKW-102-7}) is indeed satisfied.

The subsurface normal \emph{speed} $w(\boldsymbol{x})$ of a shock wave $W(%
\boldsymbol{x})=t$ is given by $w(\boldsymbol{x})=|\nabla W|^{-1}$. From (%
\ref{FKW-102-8}) the subsurface normal speed is found to be 
\begin{equation}
w(\boldsymbol{x})=\dfrac{a}{\sqrt{1+a^{2}U_{y}^{2}+a^{2}U_{z}^{2}}}.
\label{FKW-102-9}
\end{equation}

The following clarifies the shock wave geometry and kinematics relevant to
equation (\ref{FKW-102-1}). First, using (\ref{FKW-102-8}), we write the
shock wave in the form $x=at-aU(y,z)$. At initial time $t=0$, the shock wave
surface is given by $x=-aU(y,z)$. Subsequently, this surface moves uniformly
in the positive $x$-direction with speed $a$ in that direction. This motion
is not generally perpendicular to the shock wave surface at all points, and (%
\ref{FKW-102-9}) shows that the speed of the shock wave in the $x$ direction
is generally greater than the surface normal speed.

Because of their contextuality, shock waves require some care in their
specification. For instance, it is not enough to define a two-dimensional
Fock subsurface in three-dimensional space and think of it as a shock wave
of discontinuity. We need to specify the direction of motion of this surface
as well, because the Fock flow $\mathcal{F}^{(+)}[F]$ defined by $F(%
\boldsymbol{x})=t$ models Fock subsurfaces moving in the opposite direction
to those belonging to the Fock flow $\mathcal{F}^{(-)}[F]$ defined by $F(%
\boldsymbol{x})=-t$.

In the real world, irreversibility is ubiquitous: a given Fock shock flow $%
\mathcal{F}^{(+)}[W]$ may be physically observable, such as an incoming
photon or neutrino shock wave sent out from some approximate point source
such as an exploding star, whilst its theoretical counterpart $\mathcal{F}%
^{(-)}[W]$ represents an incoming sphere of radiation that would never be
seen naturally. This reinforces our earlier comments that shock waves are
essentially emergent phenomena.

\section{ Application to Maxwell's equations in vacuo}

The Fock-Kemmer analysis can be extended naturally to electromagnetic wave
theory. In this section we consider the situation of free charges in vacuo.
We shall treat the critical case of electromagnetic shock wave propagation
in a polarizable and magnetizable medium in a later section.

In vacuo, Maxwell's equations can be written in the form 
\begin{equation}
\begin{array}{rlrc}
\nabla \times \boldsymbol{B}-\dot{\boldsymbol{E}}\underset{c}{=} & 
\boldsymbol{j}_{f}, & \ \ \ \ \ \nabla \boldsymbol{\cdot }\boldsymbol{E}%
\underset{c}{=} & \rho _{f}, \\ 
\nabla \times \boldsymbol{E}+\dot{\boldsymbol{B}}\underset{c}{=} & 
\boldsymbol{0}, & \nabla \boldsymbol{\cdot }\boldsymbol{B}\underset{c}{=} & 
0,%
\end{array}
\label{FKW-103-01}
\end{equation}%
where $\dot{\boldsymbol{E}}\equiv \partial \boldsymbol{E}/\partial t$, etc,
and $\boldsymbol{j}_{f}$ and $\rho _{f}$ are the free charge current and
charge densities respectively. To deal with the fact that these equations
lead to second order wave equations, we are led to define the six component
Maxwell \emph{bi-field }$\Phi $, the \emph{bi-current density} $\boldsymbol{J%
}$, and the \emph{bi-charge density} $\Omega $ by%
\begin{equation}
\Phi \equiv 
\begin{bmatrix}
\boldsymbol{E} \\ 
\boldsymbol{B}%
\end{bmatrix}%
,\ \ \ J\equiv 
\begin{bmatrix}
\boldsymbol{j}_{f} \\ 
\boldsymbol{0}%
\end{bmatrix}%
,\ \ \ \Omega \equiv 
\begin{pmatrix}
\rho _{f} \\ 
0%
\end{pmatrix}%
,
\end{equation}%
noting that the components of $\Phi $ and $J$ are vectorial while those of $%
\Omega $ are scalar in nature. Six-component fields such as $\Phi $ and $J$
are denoted with (square) brackets whilst two-component fields such as $%
\Omega $ are denoted with (round) parentheses.

We define derivatives as 
\begin{equation}
\nabla \times \Phi \equiv 
\begin{bmatrix}
\nabla \times \boldsymbol{E} \\ 
\nabla \times \boldsymbol{B}%
\end{bmatrix}%
,\ \ \ \ \ \nabla \boldsymbol{\cdot }\Phi \equiv 
\begin{pmatrix}
\nabla \boldsymbol{\cdot }\boldsymbol{E} \\ 
\nabla \boldsymbol{\cdot }\boldsymbol{B}%
\end{pmatrix}%
,\ \ \ \ \ \dot{\Phi}\equiv 
\begin{bmatrix}
\dot{\boldsymbol{E}} \\ 
\dot{\boldsymbol{B}}%
\end{bmatrix}%
,
\end{equation}%
noting that the `divergence operator' acting on a six-component field
returns a two-component field. Then Maxwell's equations (\ref{FKW-103-01})
can be written in the form%
\begin{equation}
\dot{\Phi}+S\nabla \times \Phi \underset{c}{=}-J,\ \ \ \ \nabla \boldsymbol{%
\cdot }\Phi \ \underset{c}{=}\Omega ,  \label{FKW-103-02}
\end{equation}%
where 
\begin{equation}
S\equiv 
\begin{bmatrix}
0_{3} & -I_{3} \\ 
I_{3} & 0_{3}%
\end{bmatrix}%
,
\end{equation}%
$I_{3}$ being the $3\times 3$ identity matrix and $O_{3}$ the $3\times 3$
zero matrix. Equations (\ref{FKW-103-02}) give%
\begin{eqnarray}
\nabla \boldsymbol{\cdot }(\dot{\Phi}+S\nabla \times \Phi ) &=&\nabla 
\boldsymbol{\cdot }\dot{\Phi}=-\nabla \boldsymbol{\cdot }J,  \notag \\
\nabla \boldsymbol{\cdot }\dot{\Phi} &=&\dot{\Omega},
\end{eqnarray}%
and so we must have $\nabla \boldsymbol{\cdot }J+\dot{\Omega}=0$, which is
equivalent to the charge continuity equation $\partial _{t}\rho _{f}+\nabla 
\boldsymbol{\cdot }\boldsymbol{j}_{f}=0$.

\subsection{The Kemmer bracket of the Maxwell bi-field}

Taking the Kemmer bracket across (\ref{FKW-103-02}), we have%
\begin{equation}
\lbrack \dot{\Phi}]^{F}+S[\nabla \times \Phi ]^{F}=-[J]^{F},\ \ \ \ [\nabla 
\boldsymbol{\cdot} \Phi ]^{F}=[\Omega ]^{F}.  \label{FKW-103-03}
\end{equation}%
Applying the Fock-Kemmer identity $\lbrack \nabla \times \Phi ]^{F}=\nabla
\times \lbrack \Phi ]^{F}-\nabla F\times \lbrack \dot{\Phi}]^{F}$ to the
first equation in (\ref{FKW-103-03}) then gives 
\begin{equation}
\left\{ I_{6}-S\boldsymbol{\nabla }F\times \right\} [\dot{\Phi}%
]^{F}=-S\nabla \times \lbrack \Phi ]^{F}-[J]^{F}.\   \label{FKW-103-04}
\end{equation}%
where $I_{6}$ is the $6\times 6$ identity matrix and $S\boldsymbol{\nabla }%
F\times $ is a $6\times 6$ antisymmetric matrix with components linearly
dependent on the components of $\boldsymbol{\nabla }F$.

We now apply the Fock argument. In (\ref{FKW-103-04}), we should be able to
know everything on the right-hand side, even in the case of a shock wave, $F
= W$. This would then allow us to determine $[\dot{\Phi}]^{W}$, which is
forbidden by Fock's argument, unless the shock wave function $W$ satisfies
the condition%
\begin{equation}
\det \left\{ I_{6}-S\boldsymbol{\nabla }W\times \right\} =0.
\end{equation}%
We readily find 
\begin{equation}
\left( \nabla W\right) ^{2}=1,  \label{FKW-SHOCK}
\end{equation}%
which is precisely what we expect from special relativity ($c=1$ in this
section). We note that this condition is independent of any electric charges
in the system.

\section{The charged Dirac equation}

The charged Dirac equation in external electromagnetic fields is given by%
\begin{equation}
i\gamma ^{\mu }D_{\mu }\psi -m\psi \underset{c}{=}0  \label{FKW-110-01}
\end{equation}%
in standard notation, where the $\gamma ^{\mu }$ are the Dirac matrices \citep%
{DIRAC-1958} and $D_{\mu }\equiv \partial _{\mu }+ieA_{\mu }$ are the gauge
covariant derivatives, with $A_{\mu }$ the components of the electromagnetic
potential one-form. Separating spatial and temporal components and taking
Kemmer brackets with respect to a Fock flow $\mathcal{F}[F]$ gives%
\begin{equation}
\gamma ^{0}[D_{0}\psi ]^{F}\underset{c}{=}-\gamma ^{i}[D_{i}\psi
]^{F}-im[\psi ]^{F}.  \label{FKW-DIRAC}
\end{equation}%
Using the Fock-Kemmer identity (\ref{FKW-ID}) we can show that%
\begin{equation}
\lbrack D_{j}\psi ]^{F}=\partial _{j}[\psi ]^{F}-\partial _{j}F[D_{0}\psi
]^{F}+ie[(\partial _{j}FA_{0}+A_{j})\psi ]^{F}.
\end{equation}%
Using this in (\ref{FKW-DIRAC}) then gives%
\begin{equation}
\begin{array}{cl}
(\gamma ^{0}-\gamma ^{i}\partial _{i}F)[D_{0}\psi ]^{F}\underset{c}{=} & 
-ie\gamma ^{j}\partial _{j}[\psi ]^{F} \\ 
& -ie\gamma ^{j}[(\partial_{j}FA_{0}+A_{j})\psi ]^{F} \\ 
& -im[\psi ]^{F}.%
\end{array}%
\end{equation}%
On a shock wave, Fock's argument then gives the wavefront condition%
\begin{equation}
\det \left\{ \gamma ^{0}-\gamma ^{i}\partial _{i}W\right\} =0.
\label{FKW-D1}
\end{equation}%
Using the standard representation of the Dirac matrices \citep%
{BJORKEN+DRELL:1965B}, (\ref{FKW-D1}) then gives (\ref{FKW-SHOCK}), that is,
exactly the same condition as that found for the electromagnetic field. Note
that a veto on knowing $[D_{0}\psi ]^{W}$ is equivalent to a veto on knowing 
$[\dot{\psi}]^{W}$, since we assume we can always determine $[A_{0}\psi
]^{W} $.

\ 

We comment here on the conventional observation that Dirac quantum fields do
not commute at spacelike separations, leading to concern regarding
causality. We have four points to make about this concern.

\ 

\noindent\textbf{1)} The conventional view is that Dirac fields are not
observables but certain bilinear combinations of them are, such as the
charge four-current operator, and these observables do commute at spacelike
separations.

\ 

\noindent\textbf{2)} The Jordan-Wigner construction of fermion fields \citep%
{BJORKEN+DRELL:1965B} is manifestly non-local, supporting the view of
Schwinger that `\emph{The mathematical machinery of quantum mechanics is a
symbolic expression of the laws of atomic measurement, abstracted from the
specific properties of individual techniques of measurement.'} \citep%
{SCHWINGER-1958}. This means that with fermions, there is implicit
contextuality that may induce superluminal correlations. That does not imply
superluminal signalling.

\ 

\noindent\textbf{3)} The SPJ function for the free Dirac field does indeed
have a lightcone cutoff \citep{BJORKEN+DRELL:1965B}, which guarantees no
superluminal transmission of free field shock waves.

\ 

\noindent\textbf{4)} Somewhat surprisingly and perhaps disturbing, the
conventional Feynman propagator does not have a lightcone cutoff. However,
that propagator is used in conventional LSZ formalism scattering
calculations \citep{BJORKEN+DRELL:1965B} based on remote past (limit of time
tending to $-\infty $) \emph{in }states propagating to remote future (limit
of time tending to $+\infty $) \emph{out }states. In between state
preparation and outcome detection is the regime we call the \emph{%
information void}, where no signal detection takes place. In this regime and
in the absence of any signal detection, standard causality rules do not
apply. The rules of quantum mechanical path integrals allow (indeed require)
all dynamically possible intermediate processes to be taken into
consideration, including acausal ones. The only thing that matters
empirically is what the signal detectors register, not the imagined
behaviour of the fields in the information void.

We return to this point in \S \ref{DISTFIELDAPP}.

\section{The free Klein-Gordon equation}

As a second-order differential equation, the free particle Klein-Gordon
equation (KGE) 
\begin{equation}
\ddot{\varphi}-\nabla ^{2}\varphi +m^{2}\varphi \underset{c}{=}0
\label{FKW-987}
\end{equation}%
presents an addition layer of structure that can be circumvented by suitable
redefinition of variables. We take our cues from three places: i) Petiau 
\citep{PETIAU-1936}, Duffin \citep{DUFFIN-1938} and Kemmer \citep{KEMMER-1939}
discussed linearized approaches to the KGE along the lines of the Dirac
equation; ii) the Dirac equation can be readily discussed in Kemmer bracket
terms, and iii) the electromagnetic fields obey second order differential
equations, but our linearization approach above gave us the required shock
wave condition straightforwardly.

Our approach in this section is to introduce four extra auxiliary variables: 
$\sigma \equiv \dot{\varphi}$ and $\eta _{i}\equiv \partial _{i}\varphi $, $%
i=1,2,3$, and define the five-component field $\Phi $ by%
\begin{equation}
\Phi ^{T}\equiv (\varphi ,\sigma ,\eta _{1},\eta _{2},\eta _{3}),
\end{equation}%
where superscript $T$ denotes transpose. Then (\ref{FKW-987}) can be written
in the form%
\begin{equation}
\dot{\Phi}_{A}\underset{c}{=}\Gamma _{AB}^{i}\partial _{i}\Phi
_{B}+K_{AB}\Phi _{B},  \label{FKW-7768}
\end{equation}%
where capital Latin indices run from 1 to 5, small Latin indices run from 1
to 3, and the constant matrices $\Gamma _{AB}^{i},K_{AB}$ can be readily
determined from (\ref{FKW-987}) and (\ref{FKW-7768}). We find for example 
\begin{equation}
\Gamma ^{1}=%
\begin{bmatrix}
0 & 0 & 0 & 0 & 0 \\ 
0 & 0 & 1 & 0 & 0 \\ 
0 & 1 & 0 & 0 & 0 \\ 
0 & 0 & 0 & 0 & 0 \\ 
0 & 0 & 0 & 0 & 0%
\end{bmatrix}%
,\ \text{etc., \ \ \ and}\ \ \ K=%
\begin{bmatrix}
0 & 1 & 0 & 0 & 0 \\ 
-m^{2} & 0 & 0 & 0 & 0 \\ 
0 & 0 & 0 & 0 & 0 \\ 
0 & 0 & 0 & 0 & 0 \\ 
0 & 0 & 0 & 0 & 0%
\end{bmatrix}%
.
\end{equation}%
We note that in order to recover the original KGE (\ref{FKW-987}) from (\ref%
{FKW-7768}) we need the auxiliary equation $\boldsymbol{\eta }=\nabla
\varphi $.

The Fock shock wave condition for the KGE is found as before. Taking Kemmer
brackets on both sides of (\ref{FKW-7768}) with respect to an arbitrary Fock
flow $\mathcal{F}[F]$ and applying the Fock-Kemmer identity, we readily
deduce that 
\begin{equation}
(I_{5}+\partial _{i}F\ \Gamma ^{i})[\dot{\Phi}]^{F} \underset{c}{=} \Gamma
^{i}\partial _{i}[\Phi ]^{F}+K[\Phi ]^{F},
\end{equation}%
where $I_{5}$ is the $5\times 5$ identity matrix. On a shock wave $F=W,$
Fock's argument then leads to the condition 
\begin{equation}
\det (\partial _{i}W\Gamma ^{i}+I_{5})=0,  \label{FKW-321}
\end{equation}%
or else we would be able to determine $[\dot{\Phi}]^{W}$ from a knowledge of 
$[\Phi ]^{W}$. Condition (\ref{FKW-321}) gives $(\nabla W)^{2}=1$, which is
exactly the same as that found for the Dirac and Maxwell fields.

\section{The charged Klein-Gordon equation}

The wave equation for a charged scalar particle in external electromagnetic
potentials is given by%
\begin{equation}
D_{\mu }D^{\mu }\varphi +m^{2}\varphi \underset{c}{=}0,
\end{equation}%
where $D_{\mu }\equiv \partial _{\mu }+ieA_{\mu }$ is the same gauge
covariant operator used in the charged Dirac equation discussed above.

Our approach is a synthesis of the methods used for the charged Dirac
equation and the free Klein-Gordon equation above. First we define the fields%
\begin{equation}
\sigma \equiv D_{0}\varphi =\left( \partial _{t}+ieA_{0}\right) \varphi ,\ \
\ \ \ \eta _{i}\equiv D_{i}\varphi ,\ \ \ i=1,2,3.
\end{equation}%
Then we find%
\begin{equation}
D_{0}\eta _{i}=ieE^{i}\varphi +D_{i}\sigma ,
\end{equation}%
where $E^{i}\equiv \partial _{t}A_{i}-\partial _{i}A_{0}$ is the electric
field.

Next, we define the five component object $\Phi $ as before, that is 
\begin{equation}
\Phi ^{T}\equiv (\varphi ,\sigma ,\eta _{1},\eta _{2},\eta _{3}).
\end{equation}%
Then $\Phi $ satisfies the equation%
\begin{equation}
D_{0}\Phi _{A}\underset{c}{=}\Gamma _{AB}^{i}D_{i}\Phi _{B}+\tilde{K}%
_{AB}\Phi _{B},
\end{equation}%
where the $\Gamma ^{i}$ are as before but 
\begin{equation}
\tilde{K}=%
\begin{bmatrix}
0 & 1 & 0 & 0 & 0 \\ 
-m^{2} & 0 & 0 & 0 & 0 \\ 
ieE^{1} & 0 & 0 & 0 & 0 \\ 
ieE^{2} & 0 & 0 & 0 & 0 \\ 
ieE^{3} & 0 & 0 & 0 & 0%
\end{bmatrix}%
.
\end{equation}%
Taking Kemmer brackets with respect to a Fock flow $F$ we arrive at the
relation%
\begin{equation}
\begin{array}{cl}
(I_{5}+\Gamma ^{i}\partial _{i}F)[D_{0}\Phi ]^{F}\underset{c}{=} & \Gamma
^{j} \partial _{j}[\Phi ]^{F} +[\tilde{K}\Phi ]^{F} \\ 
& +ie\Gamma ^{j} [(\partial _{j}FA_{0}+A_{j})\Phi ]^{F} .%
\end{array}%
\end{equation}%
Applying Fock's argument leads to the same condition $\det (I_{5}+\Gamma
^{i}\partial _{i}W)=0$ as for the free Klein-Gordon field, consistent with
the expected lightcone condition $(\nabla W)^{2}=1$.

\section{The Rarita-Schwinger equation}

The success of the Standard Model is based on spin zero, spin half, and spin
one fields. Particles associated with each such spin have been observed. The
Rarita-Schwinger equation (RSE) was proposed as a model for spin
three-halves particle fields \citep{RARITA-SCHWINGER-1941}. Such a field will
have the vectorial characteristics of a spin one field and the spinorial
characteristics of a spin half field. We shall denote such a field by $\psi
_{\mu }$, the spinorial index being understood.

In line with our comments on the Dirac equation above, we should expect an
RS field not to be an observable per se. However, we did find that Dirac
field and electromagnetic field shock waves obey special relativistic
causality rules, so it is natural to see what happens in the case of RS
fields. We shall look at the free RSE, on the grounds that if that gives
superluminal shock waves, then we should not be surprised to find no stable
RS particles in nature.

With a lack of empirical evidence to guide us in choice of equation for the
RSE, we choose to work with the following RSE equation \citep%
{GASIOROWICZ-1968}:%
\begin{equation}
(i\gamma ^{\nu }\partial _{\nu }-m)\psi _{\mu }\underset{c}{=}0,\ \ \ \mu
=0,1,2,3,  \label{FKW-RSE1}
\end{equation}%
supplemented by the constraint equations 
\begin{equation}
\gamma ^{\mu }\psi _{\mu }=0,\ \ \ \partial ^{\mu }\psi _{\mu }=0.
\label{FKW-RSE}
\end{equation}%
If we did not have these constraint equations, then we could use the same
approach as we applied to the Dirac equation above to prove immediately that
the RSE field does indeed satisfy lightcone causality. Taking the Kemmer
bracket across equation (\ref{FKW-RSE}) gives%
\begin{equation}
\gamma ^{0}[\dot{\psi}_{\mu }]^{F}+\gamma ^{i}[\partial _{i}\psi _{\mu }]^{F}%
\underset{c}{=}-im[\psi _{\mu }]^{F}.  \label{FKW-001}
\end{equation}%
The Fock-Kemmer identity applied to $\psi _{\mu }$ and then used in (\ref%
{FKW-001}) gives%
\begin{equation}
\left\{ \gamma ^{0}-\gamma ^{i}\partial _{i}F\right\} [\dot{\psi}_{\mu }]^{F}%
\underset{c}{=}-\gamma ^{i}\partial _{i}[\psi _{\mu }]^{F}-im[\psi _{\mu
}]^{F}.
\end{equation}%
The Fock argument then gives us condition (\ref{FKW-D1}) for a shock wave,
exactly as for the Dirac equation.

However, this does not prove that such shock waves can be constructed: the
constraints (\ref{FKW-RSE}) may make this impossible. Our resolution of the
causality issues with the RSE equation is therefore the statement that 
\textbf{if} shock waves occurred with such fields, lightcone causality would
necessarily be maintained. That does not prove that such shock waves could
be constructed consistent with the constraints. Whatever the possibility of
such construction, superluminal propagation of spin $3/2$ particles is ruled
out.

\section{Distributional field approach}

\label{DISTFIELDAPP} Anticipating our discussion below on
electromagnetically polarizable and magnetizable media and motivated by a
desire to see the Fock analysis in more than heuristic terms, we introduce
an approach based on Fock's ideas, but now explicitly incorporating the
theory of distributions (generalized functions) and test functions. We give
a brief review of relevant distribution concepts and our notation in the
Appendix, \S \ref{App}.

Because of its discontinuity and singularity structure, a shock wave is best
not regarded as a smooth function but as a distributional-valued field.
Doing this gives some mathematical justification for Fock's argument. We
shall apply distribution methods to several situations, the first being to
revisit the first order equation discussed in \S \ref{FIRSTORDER}.

\subsection{First order equation revisited}

Given equation (\ref{FKW-102-1}), we make the shock wave ansatz 
\begin{equation}
\phi \underset{D}{=}f\theta _{W}+g\delta _{W},
\end{equation}%
where $\underset{D}{=}$ denotes \emph{distributional equality}, discussed in 
\S \ref{App}. More explicitly, we take $\phi $ to be a distribution-valued
field of the form 
\begin{equation}
\phi (t,\boldsymbol{x})\underset{D}{=}f(t,\boldsymbol{x})\theta (t-W(%
\boldsymbol{x}))+g(t,\boldsymbol{x})\delta (t-W(\boldsymbol{x}),
\end{equation}%
where $f$ and $g$ are test functions, $W$ is a Fock shock wave function, and 
$[g]^{W}\neq 0$. Then we find the derivatives%
\begin{equation}
\begin{array}{rl}
\dot{\phi}\underset{D}{=} & \dot{f}\theta _{W}+(f+\dot{g})\delta
_{W}+g\delta _{W}^{[1]}, \\ 
\boldsymbol{a}\boldsymbol{\cdot }\nabla \phi \underset{D}{=} & \boldsymbol{a}%
\boldsymbol{\cdot }\nabla f\theta _{W}+(\boldsymbol{a}\boldsymbol{\cdot }%
\nabla g-f\boldsymbol{a}\boldsymbol{\cdot }\nabla W)\delta _{W} \\ 
& -g\boldsymbol{a}\boldsymbol{\cdot }\nabla W\delta _{W}^{[1]}.%
\end{array}%
\end{equation}%
Using these expressions in the distribution field equation%
\begin{equation}
i\dot{\phi}+i\boldsymbol{a}\boldsymbol{\cdot }\nabla \phi -m\phi \underset{D}%
{=}0,
\end{equation}%
we find a distributional equation of the form 
\begin{equation}
A\theta _{W}+B\delta _{W}+C\delta _{W}^{[1]}\underset{D}{=}0,
\label{FKW-RES}
\end{equation}%
where the coefficients $A$, $B$, and $C$ are given by%
\begin{eqnarray}
A &\equiv &i\dot{f}+i\boldsymbol{a}\boldsymbol{\cdot }\nabla f-mf,  \notag \\
B &\equiv &if+i\dot{g}-if\boldsymbol{a}\boldsymbol{\cdot }\nabla W+i%
\boldsymbol{a}\boldsymbol{\cdot }\nabla g-mg,  \notag \\
C &\equiv &ig-ig\boldsymbol{a}\boldsymbol{\cdot }\nabla W.
\end{eqnarray}%
Now applying the distributional independence theorem (\ref{FKW-DISTTHEOREM})
to (\ref{FKW-RES}) we must have%
\begin{eqnarray}
A(t,\boldsymbol{x}) &=&0,\ \ \ \ \ t>W(\boldsymbol{x}),  \notag \\
\lbrack B]^{W} &=&[C^{[1]}]^{W},  \notag \\
\lbrack C]^{W} &=&0.  \label{FKW-BALANCE}
\end{eqnarray}%
It is straightforward now to show that these conditions are equivalent to $%
\boldsymbol{a}\boldsymbol{\cdot }\nabla W=1$ and $\boldsymbol{a}\boldsymbol{%
\cdot }\nabla \lbrack g]^{W}+im[g]^{W}=0$, precisely agreeing with the
results derived above using Fock's argument. We note that inside the region $%
t>W(\boldsymbol{x})$, which contains all events \emph{after} the shock wave
has passed, the field $\phi $ is essentially given by the test function $f$,
which satisfies the original wave equation (\ref{FKW-102-1}) and has no
singularities or discontinuities.

\subsection{Lorentzian signature propagation}

Suppose $\varphi $ is any field satisfying the distributional equivalence
equation of motion%
\begin{equation}
\square _{v}\varphi \equiv v^{-2}\partial _{t}^{2}{\varphi }-\nabla
^{2}\varphi \underset{D}{=}V(\partial _{\mu }\varphi ,\varphi ,\ldots ),
\label{FKW-Properties}
\end{equation}%
where $v$ is a constant and the highest derivative on the right hand side is
first order in time. Consider the shock wave ansatz%
\begin{equation}
\varphi =f\theta _{W}+g_{0}\delta _{W}+g_{1}\delta _{W}^{[1]}+\ldots
g_{n}\delta _{W}^{[n]},  \label{FKW-phi}
\end{equation}%
for some finite integer $n\geqslant 0$, with $\theta _{W}\equiv \theta (t-W(%
\boldsymbol{x}))$, $\delta _{W}\equiv \delta (t-W(\boldsymbol{x}))$, and $f$
and the $\{g_{k}:k=0,1,\ldots ,n\}$ are a set of test functions with $%
[g_{n}]^{W}\neq 0$. Here $W(\boldsymbol{x})$ is some Fock precursor shock
wave function whose properties are to be determined from (\ref%
{FKW-Properties}). Then applying the distributional equivalence theorem
quoted in the Appendix, we readily conclude that%
\begin{equation}
\lbrack g_{n}]^{W}(1-v^{2}\nabla W\boldsymbol{\cdot }\nabla W)=0,
\end{equation}%
plus other conditions not relevant to the conclusions. Since we have assumed 
$[g_{n}]^{W}\neq 0$, we deduce that the shock wave precursor function $W$
must satisfy the condition $v^{2}\nabla W\boldsymbol{\cdot }\nabla W=1$.
From (\ref{FKW-speed}), the shock wave normal speed is therefore $v$.

A particular issue arises with non-linear theories, because products of
distributions are not defined here. Therefore, any terms on the right-hand
side of (\ref{FKW-Properties}) such as $\varphi ^{2}$ would in principle
create a problem with the distributional approach. Our resolution is to look
at the physics of observation. It is well-known that conventional quantum
field theory encounters renormalization divergences that are removed by an
appeal to the finiteness of observed quantities. Indeed, products of quantum
field operators are generally ill-defined. In our case, we would argue that
non-linear interaction terms, such as $\varphi ^{2}$ on the right-hand side
of (\ref{FKW-Properties}) should be re-interpreted, because shock waves are
the \emph{results} of field interactions. For example, we would propose an
ansatz for $\varphi ^{2}$ of the form%
\begin{equation}
\varphi ^{2}\sim f^{2}\theta _{W}+G_{0}\delta _{W}+G_{1}\delta
_{W}^{[1]}+\ldots G_{n}\delta _{W}^{[n]},
\end{equation}%
where $f$ is the same test function as in equation (\ref{FKW-phi}) and the $%
G_{i}$ are test functions. Such an ansatz does not then alter our
conclusions. This argument is analogous to Wilson's expansion of products of
quantum fields \citep{WILSON-1969}. We note that in our discussion of
polarizable and magnetizable media in the next section, we take a similar
approach in our modelling of the polarization and magnetization \emph{%
response} to an incoming electromagnetic shock wave.

The same methodology allows us to deduce the same result for any higher
order equation such as%
\begin{equation}
\alpha (\square _{v})^{2}\varphi +\beta \square _{v}\varphi \underset{D}{=}%
V(\partial _{\mu }\varphi ,\varphi ,\ldots ),
\end{equation}%
where $\alpha $ and $\beta $ are test functions.

\subsection{Shock waves in polarizable and magnetizable media}

The possibility that $v$ is not the speed of light in vacuo ($c=1$
throughout this paper) is of critical importance in the theory of
propagation of electromagnetic waves through real media. It is possible, in
certain cases of anomalous dispersion, to encounter situations where $v>1$.
We discuss what must happen in such cases in this section.

The modified d'Alembertian operator $\square _{v}\equiv v^{-2}\partial
_{t}^{2}-\nabla ^{2}$ is the critical factor in any discussion of wave
processes. In electromagnetic wave theory the constant $v$ is referred to as
the \emph{phase velocity.} It is usually asserted that light propagates in a
polarizable and/or magnetizable medium with this speed. When $v$ is less
than $c$, the speed of light in vacuo, it is possible for particle speeds in
such a medium to exceed $v$ (but still be less than $c$), and then Cerenkov
or Askaryan radiation may be observed. These are all important observed
phenomena. The problem however is that it is possible to encounter media for
which $v$ exceeds $c$, as well as media in which the group velocity is
greater than $c$. In such cases, the obvious question is whether precursor
signals could ever propagate faster than $c$.

It is not enough to simply assert the traditional relativistic veto $v
\leqslant c$: the dynamics of light propagation in media should predict that
veto in a natural, accountable way. We discuss here how the distributional
field method deals with this issue.

In any such discussion, it is important to understand that we are dealing
with complex, emergent processes using reductionist equations of motion.
Therefore, approximate, relatively simple models have to made, generally
regarded as statistical in nature. Maxwell's equations for electromagnetic
fields in polarizable and magnetizable media are exactly of this type \cite%
{STRATTON-1941}. We shall apply our distributional field method to
electromagnetic waves in a nominally linearly polarizable and magnetizable
homogeneous, isotropic medium, with charge-free field equations of motion%
\begin{equation}
\begin{array}{cc}
\nabla \boldsymbol{\cdot }\boldsymbol{B}=0, & \nabla \times \boldsymbol{E}+\partial
_{t}\boldsymbol{B}=\boldsymbol{0}, \\ 
\nabla \boldsymbol{\cdot }\boldsymbol{D}\underset{c}{=}0, & \nabla \times 
\boldsymbol{H}-\partial _{t}\boldsymbol{D}\underset{c}{=}\boldsymbol{0}.%
\end{array}
\label{FKW-MAXWELL}
\end{equation}%
Here $\boldsymbol{D}\equiv \epsilon _{0}\boldsymbol{E}+\boldsymbol{P}$ is the \emph{%
displacement} field, where $\epsilon _{0}$ is the permittivity of free space
and $\boldsymbol{P}$ is the \emph{polarization} field, and $\boldsymbol{H}\equiv 
\boldsymbol{B}/\mu _{0}-\boldsymbol{M}$ is the \emph{magnetic intensity }field,
where $\mu _{0}$ is the permeability of free space and $\boldsymbol{M}$ is the 
\emph{magnetization }field. Equations (\ref{FKW-MAXWELL}) are generally
taken as exact equations, within the given context.

For linear, isotropic media, the polarization and magnetization fields are
generally assumed to be given by 
\begin{equation}
\boldsymbol{P}=\chi _{e}\epsilon _{0}\boldsymbol{E},\ \ \ \ \ \boldsymbol{M}=\chi _{m}%
\boldsymbol{H},
\end{equation}
where $\chi _{e}$ is the \emph{electric susceptibility} and $\chi _{m}$ is
the \emph{magnetic susceptibility}. Assuming these susceptibilities are
scalar constants, then equations (\ref{FKW-MAXWELL}) give%
\begin{equation}
(\varepsilon \mu \partial _{t}^{2}-\nabla ^{2})\boldsymbol{E}\underset{c}{=}%
\boldsymbol{0},\ \ \ (\varepsilon \mu \partial _{t}^{2}-\nabla ^{2})\boldsymbol{B}%
\underset{c}{=}\boldsymbol{0}\text{,}  \label{FKW-ANOMALOUS}
\end{equation}%
where $\epsilon \equiv (1+\chi _{e})\epsilon _{0}$,\ $\mu \equiv (1+\chi
_{m})\mu _{0}$, and so the phase velocity $v$ is given by $v\equiv 1/\sqrt{%
\varepsilon \mu }$. The significance to us here is that in the case of
certain novel media, it is possible to encounter \emph{negative}
susceptibilities, leading to the result $v>c$. In such cases, our above
discussion of Lorentzian signature propagation leads us to conclude that
equations (\ref{FKW-ANOMALOUS}) must be incorrect equations for precursor
wavefront propagation. We resolve this problem in two steps.

First, making no linearity assumption about polarization or magnetization,
equations (\ref{FKW-MAXWELL}) give the exact wave equations%
\begin{equation}
\begin{array}{rl}
(\epsilon _{0}\mu _{0}\partial _{t}^{2}-\nabla ^{2})\boldsymbol{E}\underset{c}{=}
& -\epsilon _{0}^{-1}\nabla (\nabla \boldsymbol{\cdot }\boldsymbol{P})-\mu
_{0}\partial _{t}^{2}\boldsymbol{P} \\ 
& \ \ \ \ \ \ \ \ \ \ \ \ \ \ \ -\mu _{0}\nabla \times \partial _{t}\boldsymbol{M%
}, \\ 
(\epsilon _{0}\mu _{0}\partial _{t}^{2}-\nabla ^{2})\boldsymbol{B}\underset{c}{=}
& \mu _{0}\nabla \times \partial _{t}\boldsymbol{P}+\mu _{0}\nabla \times
(\nabla \times \boldsymbol{M}). \label{FKW-RESPONSE}%
\end{array}%
\end{equation}

Second, we invoke causality. Suppose we have a nominally linear, isotropic
medium at rest in the laboratory \emph{before }any shock wave has passed
through. Then clearly, all fields are zero then. Now suppose a precursor
shock wave passes through the medium. The medium will consist of atoms and
molecules that cannot react instantly. There must be some delay before the
polarisation $\boldsymbol{P}$ and magnetization $\boldsymbol{M}$ can adjust to the
sudden changes in the electric and magnetic fields $\boldsymbol{E}$ and $\boldsymbol{%
B}$. Therefore, we make the following shock wave ansatz for the fields
concerned, all of which are regarded now as distributional fields:%
\begin{equation}
\begin{array}{cl}
\boldsymbol{E}\underset{D}{=}\overline{\boldsymbol{E}}\theta _{W}+\displaystyle{%
\sum_{n=0}^{N}}\boldsymbol{E}_{n}\delta _{W}^{[n]}, & \boldsymbol{P}\underset{D}{=}%
\chi _{e}\varepsilon _{0}\overline{\boldsymbol{E}}\theta _{W}+\displaystyle{%
\sum_{n=0}^{N-1}}\boldsymbol{P}_{n}\delta _{W}^{[n]}, \\ 
\boldsymbol{B}\underset{D}{=}\overline{\boldsymbol{B}}\theta _{W}+\displaystyle{%
\sum_{n=0}^{N}}\boldsymbol{B}_{n}\delta _{W}^{[n]}, & \boldsymbol{M}\underset{D}{=}%
\displaystyle{\frac{\chi _{m}}{\mu }}\overline{\boldsymbol{B}}\theta _{W}+%
\displaystyle{\sum_{n=0}^{N-1}}\boldsymbol{M}_{n}\delta _{W}^{[n]}, \\ 
& [\boldsymbol{E}_{N}]^{W},[\boldsymbol{B}_{N}]^{W}\neq \boldsymbol{0},%
\end{array}
\label{FKW-ANSATZ}
\end{equation}%
for some integer $N\geqslant 1$, noting the different upper limits on the
summations in $\boldsymbol{E}$ and $\boldsymbol{B}$ compared to those in $\boldsymbol{P}$
and $\boldsymbol{M}$. This difference is our method of encoding causality. In
these expressions, the coefficients of $\theta _{W}$ and the $\delta
_{W}^{[n]}$ are assumed to be test function fields.

Applying the distributional equivalence theorem to equations (\ref%
{FKW-RESPONSE}) now considered as distributional field equations when ansatz
(\ref{FKW-ANSATZ}) is used, there are two important conclusions. On the one
hand, the coefficients of $\theta _{W}$ give the wave equations%
\begin{equation}
(\epsilon \mu \partial _{t}^{2}-\nabla ^{2})\overline{\boldsymbol{E}}=\boldsymbol{0}%
,\ \ \ (\epsilon \mu \partial _{t}^{2}-\nabla ^{2})\overline{\boldsymbol{B}}=%
\boldsymbol{0}
\end{equation}%
in the region of the medium where $t>W(\boldsymbol{x})$, that is, after the
precursor shock wave has passed. These wave equations have phase velocity $%
v=1/\sqrt{\varepsilon \mu }$, with no restriction on $v$ being greater than $%
c$. On the other hand, matching the effects of the $\delta _{W}^{[N]}$ terms
in the ansatz leads to the conditions%
\begin{equation}
\lbrack \boldsymbol{E}_{N}]^{W}(\varepsilon _{0}\mu _{0}-\nabla W \boldsymbol{%
\cdot} \nabla W)=[\boldsymbol{B}_{N}]^{W}(\varepsilon _{0}\mu _{0}-\nabla W 
\boldsymbol{\cdot} \nabla W)=0,
\end{equation}%
from which we deduce the expected shock wave condition $c^{2}\nabla W 
\boldsymbol{\cdot} \nabla W=1$. The distributional field approach therefore
allows phase speeds greater than $c$ in media \textbf{after} signals have
passed, but retains the relativistic lightcone limit on precursor signal
propagation itself.

\subsection{Shock waves in quantum field theory}

Up to this point, we have been discussing classical fields. Shock waves in
quantum field theory present new challenges, principally on account of the
uncertainty principle. If we prepare a localized-in-space signal state, then
we can expect a spread in momentum\textbf{\ }associated with that state.
Indeed, the concept of particle state in quantum field theory remains
problematical \citep{COLOSI+ROVELLI-2009}. We make two comments here,
reserving this topic for future work.

First, the SPJ function $\Delta (x)$ for the scalar field demonstrates
precisely the sort of structure that our distributional field approach has
taken. Specifically, the SPJ function for the free Klein-Gordon equation (%
\ref{FKW-987}) is given by%
\begin{equation}
\Delta (t,\boldsymbol{r})=\frac{m}{4\pi \sqrt{t^{2}-r^{2}}}J_{1}(m\sqrt{%
t^{2}-r^{2}})\theta (t-r)-\frac{1}{4\pi r}\delta (t-r),
\end{equation}%
for $t > 0$, which means that precursor shock waves are limited by the speed
of light ($c=1 $ here).

The second point concerns the Feynman propagator. It is well-known that the
scalar field propagator $\Delta _{F}(x)$ given by the famous $+i\epsilon $
prescription,%
\begin{equation}
\Delta _{F}(x)\equiv \int \frac{d^{4}p}{(2\pi )^{4}}\frac{e^{-ipx}}{%
p^{2}-m^{2}+i\epsilon }
\end{equation}%
has the merit of transmitting positive energy signals forwards in time, and
`negative energy waves backwards in time', according to the
Feynman-Stueckelberg interpretation \citep{BJORKEN+DRELL:1964A}. However, $%
\Delta _{F}(x)$ does not vanish outside the lightcone, raising the question
of precursor shock wave speeds once again. The conventional resolution is to
assert that real signals cannot be sent faster than $c$, and that whatever
is transmitted outside the lightcone via the Feynman propagator concerns 
\emph{correlations}, which are not signals. Correlations are emergent
phenomena, underling the point that quantum field theory is really a theory
of observation processes, rather than ``things'' such as fields or particles.

On the same point, it is remarkable that Julian Schwinger developed a novel
approach to quantum field theory called source theory, in which the emphasis
is on signal preparation and signal detection. In his approach, he
postulated that the vacuum-to-vacuum amplitude $Z[J] \equiv \langle
0^{+}|0^{-}\rangle ^{J}$ in the presence of sources $J$ is of the form%
\begin{equation}
Z[J] \equiv \exp [(i/2)\int d^{4}xd^{4}yJ(x)\Delta _{+}(x-y)J(y)].
\label{FKW-SCHWINGER}
\end{equation}%
Close inspection (\citep{SCHWINGER-1969}) shows that in fact, $\Delta
_{+}(x)=\Delta _{F}(x).$ Therefore, we deduce that Schwinger's approach
would \emph{not} address the precursor shock wave issue as it stands.

By this we mean the following. Suppose the apparatus \emph{creating} signals
occupied a finite region $R_{1}$ of spacetime and the apparatus \emph{%
detecting} signals occupied another finite region $R_{2}$ of spacetime such
that $R_{1}$ and $R_{2}$ are disjoint. We may write the source function $J$
as $J=J_{1}+J_{2}$, where $J_{1}$ has support in $R_{1}$ and $J_{2}$ has
support in $R_{2}$. Then Schwinger's amplitude (\ref{FKW-SCHWINGER}) can be
written as%
\begin{equation}
Z[J]\sim \exp [\frac{i}{2}\int_{R_{2}}d^{4}x\int_{R_{1}}d^{4}yJ_{2}(x)\left\{ 
\begin{array}{c}
\Delta _{+}(x-y) \\ 
+\Delta _{+}(y-x)%
\end{array}%
\right\} J_{1}(y)],  \label{FKW-SSHOCK}
\end{equation}%
ignoring the pieces where apparatus in a given region interacts with itself.

We define $L(R_{1})$ to be the `lightcone' associated with $R_{1}$, by which
we mean the set of all those events in spacetime that are each in or on the
lightcone of at least one event in $R_{1}.$ Now suppose $R_{2}$ has zero
intersection with $L(R_{1})$, which means that all events in $R_{2}$ are
spacelike relative to all points in $R_{1}$. The point is, the Feynman
propagator does not vanish between such points. Therefore, according to (\ref%
{FKW-SSHOCK}) a shock wave initiated at $R_{1}$ would have a non-zero effect
on $Z[J]$, contrary to intuition.

We note that it is an inadequate argument to dismiss this result on the
grounds that only correlations are involved, or that the effects are
`small'. There is a problem here of principle touching on the relationship
between classical relativity and quantum mechanics, and on the generally
under-developed status of the theory of localized observation in quantum
field theory.

The Feynman propagator is used conventionally because of the \emph{input}
that positive energies propagate forwards in time, but this input comes at
the cost of violating the lightcone veto. It works conventionally because of
the temporal limits to infinity being taken. Problems arise when this cannot
be done, as in the case of shock waves. Our thoughts here are that it is
possible to make an alternative choice in Schwinger's formalism that uses
the lightcone veto as an \emph{input}, at the expense of the positive energy
input. Specifically, we could make the replacement $\Delta
_{F}(x)\rightarrow \Delta _{C}(x)$, where $\Delta _{C}(x)\equiv -\frac{1}{2}%
(\Delta _{R}(x)+\Delta _{A}(x))$. The retarded and advanced propagators $%
\Delta _{R}$ and $\Delta _{A}$ satisfy the same inhomogeneous equation as $%
\Delta _{F}$ (up to a sign) but most significantly, vanish outside the
lightcone. If we did this, then shock waves initiated in region $R_{1}$
would never affect detectors in $R_{2}$ if $R_{2}$ and $L(R_{1})$ were
disjoint. We note that $\Delta _{C}(x)$ differs from $\Delta _{F\text{ }}$
only by a complementary function $\Delta _{H}(x)$.

There are two points about this suggestion. First, Schwinger aimed to avoid
fields \emph{per se} in his formalism. A replacement such as the one
suggested here would need some interpretation in terms of standard field
operators, particularly the creation and annihilation operators. We note
that what observers see in their detectors are \emph{signals}, not
necessarily positive energy particles. Second, Schwinger did actually
consider the possibility of adding complementary functions into his
formalism \citep{SCHWINGER-1998-1}.

\section{Concluding remarks}

Quantum field theory shock wave analysis appears not to have been
significantly explored yet. It is our belief that any development of it will
require considerable attention to the observer concept and more explicit
modelling of the processes of observation. This will require taking emergent
concepts such as irreversibility and finite time processes in quantum field
theory into account much more than they are at present.

Our distributional approach uses the most singular term in the shockwave ansatz, such as in equations (\ref{FKW-phi}) and (\ref{FKW-ANSATZ}). However, much interesting detail can be found in the less singular terms, such as information concerning
the flow of energy and momentum via shock wave fronts, particularly in the
case of electromagnetic waves. We hope to report further on those details in
subsequent articles. 

\section*{Acknowledgements}
R. H. A. is grateful for financial support from the Kurdistan regional
government (KRG). He thanks his family for
their deep patience in undertaking their son's responsibilities during his
absences. G. J. is indebted to Nicholas Kemmer for inspiring this work.

\appendix

\section{Distribution theory}

\label{App}

There are two spaces of objects in our approach, referred to as \emph{%
distributions} and \emph{test functions} respectively.

If $D$ is a distribution and $f$ a real or complex-valued test function over 
$\mathbb{R}$, the action $\langle D,f\rangle $ of $D$ on $f$ is defined by%
\begin{equation}
\langle D,f\rangle \equiv {\displaystyle{\int_{-\infty }^{\infty }}}%
D(x)f(x)dx,
\end{equation}%
and is assumed to exist for all distributions and test functions.

The value $f(a)$ at $x=a$ of a test function $f$ is denoted by $[f]^{a}$.
The $n^{th}$ derivative of a test function $f$ is also a test function and
denoted by $f^{[n]}, \ n=0,1,2, \cdots \ldots$, with $f^{[0]} \equiv f$.

\subsection*{test functions}

A \emph{test function} \citep{STREATER+WIGHTMAN-1964} is an infinitely
differentiable real or complex-valued function that falls off sufficiently
rapidly as $|x|\rightarrow \infty $ , such that

\begin{enumerate}
\item $\langle 1,f^{[n]}\rangle $ exists for $n=0,1,2,\ldots $.

\item If $f$ and $g$ are test functions, and $\alpha $ and $\beta $ are real
or complex constants, then $\alpha fg$ and $\alpha f+\beta g $ are test
functions.
\end{enumerate}

\subsection*{Distributions}

A \emph{distribution} $D$ is a process that maps a test function into $%
\mathbb{R}$ or $\mathbb{C}$ via the processes of integration, subject to the
following conditions for any test function $f$:

\begin{enumerate}
\item[3.] For any constant $\alpha $ and any test function $f$, we have $%
\langle \alpha D,f\rangle =\alpha \langle D,f\rangle$.

\item[4.] For any distributions $D_{1}$, $D_{2}$ we define their sum $%
D_{1}+D_{2}$ as $\langle \{D_{1}+D_{2}\},f\rangle \equiv \langle
D_{1},f\rangle +\langle D_{2},f\rangle $.

\item[5] For any distribution $D$, we define its $n^{th}$ derivative $%
D^{[n]} $ in terms of its action on any test function $f$ by $\langle
D^{[n]},f\rangle \equiv (-1)^{n}\langle D,f^{[n]}\rangle , \ \ \ n =
0,1,2,\cdots $.

\item[6] For any distribution $D$ and test functions $f$, $g$, we define the
generalized function $fD$ by $\langle fD,g\rangle \equiv \langle D,fg\rangle 
$.
\end{enumerate}

Important examples of distributions are

\subsection*{The Heaviside step $\protect\theta _{a}$}

The conventional notation for this distribution is $\theta _{a}(x)\equiv
\theta (x-a)$, where $a$ is real. For any test function $f$, $\theta _a$ is
defined by 
\begin{equation}
\langle \theta _{a},f\rangle \equiv \displaystyle{\int_{a}^{\infty }}f(x)dx.
\end{equation}

\subsection*{The reverse Heaviside step $\overline{\protect\theta }_{a}$}

The conventional notation for this distribution is $\overline{\theta }%
_{a}(x)\equiv \theta (a-x)$. For any test function $f$, $\overline{\theta}_a$%
is defined by 
\begin{equation}
\langle \overline{\theta }_{a},f\rangle \equiv \displaystyle{\int_{-\infty
}^{a}}f(x)dx.
\end{equation}

\subsection*{The Dirac delta $\protect\delta _{a}$}

The conventional notation for this distribution is $\delta _{a}(x)\equiv
\delta (a-x)$ or $\delta (x-a)$. For any test function, $\delta _a$ is defined
by 
\begin{equation}
\langle \delta _{a},f\rangle =[f]^{a}.
\end{equation}

\subsection*{Distributional equivalence}

Two distributions $D_{1}$, $D_{2}$ are \emph{distributionally equivalent},
written $D_{1}\underset{D}{=}D_{2}$, if $\langle D_{1},f\rangle =\langle
D_{2},f\rangle $ for any test function $f$.

Using the rules given above, a number of distributions involving Heaviside
steps and Dirac deltas can be shown to be distributionally equivalent, such
as

\begin{enumerate}
\item $\theta _{a}+\overline{\theta }_{a}\underset{D}{=}1$,

\item $\theta _{a}^{[1]}\underset{D}{=}\delta _{a}$,

\item $\overline{\theta }_{a}^{[1]}\underset{D}{=}-\delta _{a}$,

\item For the product of any test function $f$ and the Dirac delta, we can
choose to evaluate $f$ at $x=a$ or not, that is, 
\begin{equation}
f\delta _{a}\underset{D}{=} [f]^{a}\delta _{a}.
\end{equation}
Differentiating this last distributional equivalence on both sides with
respect to $x$ gives the rule 
\begin{equation}
f^{[1]}\delta _{a}\underset{D}{=}([f]^{a}-f) \delta _{a}^{[1]},
\end{equation}
and so on for higher derivatives.
\end{enumerate}

Using the above rules, we can prove the following:

\

\noindent \textbf{Theorem:} The distributions $\theta _{a}$, $\overline{%
\theta }_{a}$, $\delta _{a}$, $\delta _{a}^{[1]}$, etc., are \emph{%
distributionally independent}, which means the following. Suppose $e$, $f$, $%
g_{0},g_{1}$, $g_{2},\ldots $, are test functions and we are given that
\begin{equation}
e\theta _{a}+f\overline{\theta }_{a}+\sum_{n=0}^{\infty }g_{n}\delta
_{a}^{[n]}\underset{D}{=}0\text{,}  \label{FKW-200-01}
\end{equation}%
where $\delta _{a}^{[0]}\equiv \delta _{a}$. Then assuming we can
interchange orders of summation, we must have%
\begin{eqnarray}
e(x) &=&0,\ \ \ x>a,  \notag \\
f(x) &=&0,\ \ \ x<a,  \notag \\
0 & = & \sum_{p=0}^{\infty }(-1)^{m+p}{{m+p}\choose{p}}[g_{m+p}^{[p]}]^{a}, \notag \\
& & \ \ \ \ \  m=0,1,2,3,\ldots  \label{FKW-DISTTHEOREM} 
\end{eqnarray}

\end{document}